\documentclass[aps,showpacs,pra,twocolumn]{revtex4-1}
\usepackage[latin1]{inputenc}
\usepackage{bm}
\usepackage{multirow,amssymb,amsbsy,amsmath}
\usepackage{graphicx}
\usepackage{verbatim}
\makeatletter
\usepackage{pifont}
\makeatother

\usepackage{color}

\begin{document}



\title{Experimental detection of polarization-frequency quantum correlations
in a photonic quantum channel by local operations}

\author{Jian-Shun Tang}
\author{Yi-Tao Wang}
\author{Geng Chen}
\author{Yang Zou}
\author{Chuan-Feng Li}
\email{cfli@ustc.edu.cn}
\author{Guang-Can Guo}
\affiliation{Key Laboratory of Quantum Information, University of Science and
Technology of China, CAS, Hefei, Anhui 230026, China\\
Synergetic Innovation Center of Quantum Information $\&$ Quantum Physics,
University of Science and Technology of China, Hefei, Anhui 230026, China}

\author{Ying Yu}
\author{Mi-Feng Li}
\author{Guo-Wei Zha}
\author{Hai-Qiao Ni}
\author{Zhi-Chuan Niu}
\email{zcniu@semi.ac.cn}
\affiliation{The state key laboratory for superlattices and microstructures, Institute
of semiconductors, CAS, PO Box 912, Beijing 100083}

\author{Manuel Gessner}
\author{Heinz-Peter Breuer}
\email{breuer@physik.uni-freiburg.de}
\affiliation{Physikalisches Institut, Universit\"at Freiburg,
Hermann-Herder-Stra{\ss}e 3, D-79104 Freiburg, Germany}

\date{\today}

\begin{abstract}
The measurement of correlations between different degrees of freedom is an important, but in general extremely difficult task in many applications of quantum mechanics. Here, we report an all-optical experimental detection and quantification of quantum correlations between the polarization and the frequency degrees of freedom of single photons by means of local operations acting only on the polarization degree of freedom. These operations only require experimental control over an easily accessible two-dimensional subsystem, despite handling strongly mixed quantum states comprised of a continuum of orthogonal frequency states. Our experiment thus represents a photonic realization of a scheme for the local detection of quantum correlations in a truly infinite-dimensional continuous-variable system, which excludes an efficient finite-dimensional truncation.
\end{abstract}


\maketitle

\section{Introduction}
Photons represent natural carriers of quantum information and play a crucial role in quantum communication and quantum computation \cite{Zoller}. As flying qubits they provide the messengers among different nodes in quantum communication protocols. During transmission, however, a photonic qubit will unavoidably couple to an environment, i.e., it will become an open quantum system. On the one hand, correlations between the qubit state and its environment can significantly influence the qubit's time evolution \cite{ringbauer2015} and thereby lead to detrimental effects on the capacity of the quantum channel. System-environment correlations are responsible for non-Markovian effects \cite{TheWork,lofranco2013,breuer2015}, and, when present in the initial state, can even lead to a complete breakdown of any standard master equation description \cite{Gorini,Pechukas,Lindblad,Royer,Buzek}. On the other hand, quantum correlations form the foundation for quantum information technologies. Quantum entanglement, for instance, provides the quantum advantage for quantum key distribution \cite{ekert1991} and quantum teleportation \cite{bennett1993}, while quantum discord \cite{modi2012} plays an important role for entanglement activation \cite{piani2011,streltsov2011,Adesso2014,Orieux2015} and distribution \cite{streltsov2012}, as well as for quantum metrology \cite{modi2011,girolami2013}. It is therefore of both fundamental and practical importance to develop efficient ways to characterize quantum correlations in photonic systems.

In optical experiments, quantum correlations of small systems, e.g., a photon pair, can be detected by performing tomography on the total system \cite{xujs2010}. However, for high-dimensional systems, resource constraints and lack of control soon render such measurements impossible. A particularly complex, yet common situation is encountered when the photon's polarization state interacts with its own frequency modes, for instance in a birefringent crystal \cite{licf2011}. In this case, correlations emerge between the polarization state, which encodes the qubit, and a continuum of frequency states. This type of quantum channel is frequently used in quantum information applications even though the environmental state describing the frequency modes is not fully controllable and hard to determine.

In contrast, the $2$-dimensional subsystem, i.e., the polarization qubit, can be fully controlled and detected with local operations even in such a complex system. The question thus arises, does this allow us to extract information about the quantum correlation of the entire system? One strategy to answer this question is to use the time evolution of reduced states of the $2$-dimensional subsystem. Detecting, for instance, an increase of the trace distance between pairs of states in the subsystem constitutes a witness for total correlations in one of the initial states \cite{laine2010}. This scheme was applied experimentally in optical architectures \cite{licf2011,smirne2011}. Further theoretical extensions allow for the detection of quantum correlations of an unknown system-environment state using only local operations on the system \cite{gessner2011}, as demonstrated in a trapped-ion experiment \cite{gessner2014}. The method can also be used to discriminate between quantum and classical correlations \cite{Cialdi2014} and, quite recently, an application to the local detection of a quantum phase transition has been developed \cite{Gessner2014epl}.

In this letter we focus on the open quantum system formed by the polarization
degree of freedom of a single photon which is coupled to its environmental
frequency degrees of freedom in a birefringent quantum channel \cite{licf2011}.
Single photons are generated from a self-assembled quantum dot (QD) and initial
system-environment correlations are created by sending the photon through a
calcite crystal. We show that it is possible to experimentally detect and quantify the
quantum correlations in this system by carrying out only local operations on the
polarization degree of freedom of the photon.

\section{Theoretical framework}
First, we briefly recall the theoretical framework of the local detection method,
introduced in Ref.~\cite{gessner2011}.
Given a photon state $\rho_{SE}$, where ``$S$" represents the
polarization subsystem and ``$E$" represents the frequency modes, how can we
measure the quantum correlation of this state? One method is to map this state to
an associated state $\rho_{SE}'$, which contains only classical correlations, and
then calculate the distance between $\rho_{SE}$ and $\rho_{SE}'$. This map can
be realized through a controlled local dephasing process, i.e.,
\begin{equation} \label{dephasingmap}
\rho_{SE}\rightarrow\rho_{SE}'=\sum_{\mu}\Pi_{\mu}\rho_{SE}\Pi_{\mu},
\end{equation}
where $\Pi_{\mu}=|\mu\rangle\langle\mu|\otimes I$ and the $|\mu\rangle$ are the
eigenstates of the reduced state $\rho_{S}={\rm{Tr}}_{E}\{\rho_{SE}\}$. The
quantum discord of $\rho_{SE}'$ is zero \cite{gessner2011}. Therefore, the
quantum correlations in $\rho_{SE}$ can be quantified as \cite{luo2008,bellomo2012,aaronson2013}
\begin{equation} \label{qctheory}
 \delta(\rho_{SE}) = ||\rho_{SE}-\rho_{SE}'||,
\end{equation}
where $||A||=\rm{Tr}\sqrt{A^{\dagger}A}$ denotes the trace norm. We note
that the trace distance~(\ref{qctheory}) can be interpreted in terms of the
probability for a successful discrimination of the two quantum states by a single
measurement \cite{Hayashi}. Suppose now that both $\rho_{SE}$ and
$\rho_{SE}'$ evolve in time $\tau$ through the same unitary evolution operator
$U(\tau)$ leading to the reduced open system state
$\rho_S(\tau)={\rm{Tr}}_E\left\{ U(\tau) \rho_{SE} U^{\dagger}(\tau)\right\}$
and $\rho'_S(\tau)={\rm{Tr}}_E\left\{ U(\tau) \rho'_{SE} U^{\dagger}(\tau)\right\}$,
respectively. Since the partial trace $\rm{Tr}_E$ over the environment is
a positive operation, the local trace distance
\begin{equation}
 \Delta(\tau) = || \rho_{S}(\tau) - \rho'_{S}(\tau) ||
\end{equation}
represents a lower bound for the quantum correlations in the state $\rho_{SE}$ \cite{ruskai1994},
i.e. we have
\begin{equation} \label{ineq}
 \Delta(\tau) \leq \delta(\rho_{SE}).
\end{equation}
As this relation holds for all $\tau$ we find
\begin{equation}
 \max_{\tau} \Delta(\tau) \leq \delta(\rho_{SE}).
\end{equation}
According to this general inequality the quantity on the left-hand side -- a local quantity accessible through only local operations on the open system $S$ --
represents a witness for the quantum correlations in the total
system-environment state $\rho_{SE}$ \cite{gessner2013}.

\section{Experimental framework and results}
\begin{figure}[tb]
\centering
\includegraphics[width=0.9\linewidth]{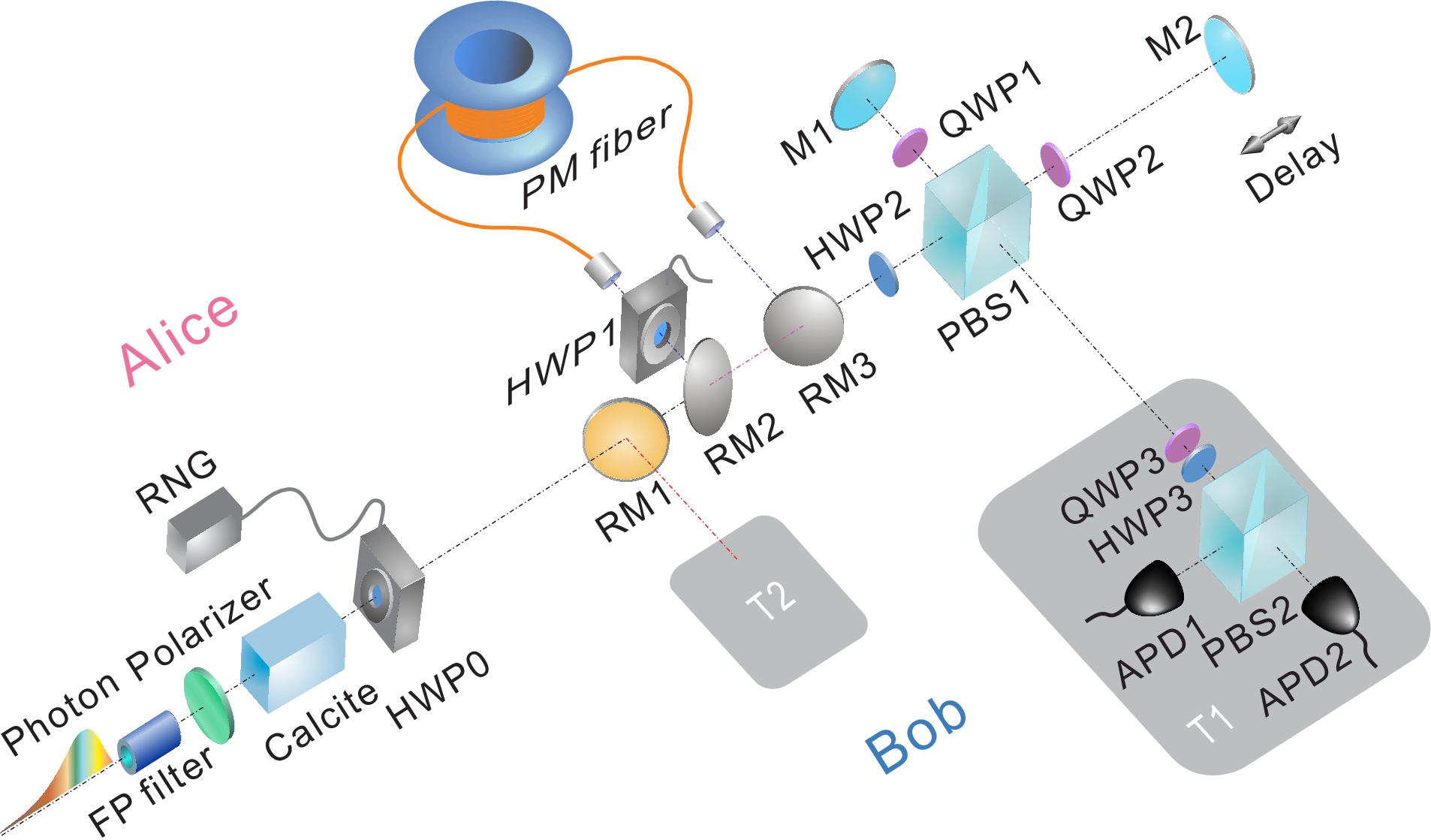}
\caption{\label{Fig1} The system is realized by the polarization state and the frequency modes of single photons. Alice generates and filters out the single photons, then sends the photons to a polarizer followed by a calcite crystal, realizing a quantum channel, and finally to a random-number-generator (RNG) driven half-wave plate (HWP0) to prepare the initial state $\rho_{SE}$ which is sent to Bob. Bob measures the quantum correlation of the initial state using local operations on the polarization state. His setup consists of four sections: (1) Three removable mirrors (RM1, RM2, RM3) direct the photons to different functional modules; (2) A computer-controlled half-wave plate (HWP1) and a long PM fiber constitute the controlled dephasing map in which the state $\rho_{SE}'$ is created; (3) A Michelson delay setup including HWP2, PBS1, QWP1, QWP2, M1 and M2 generate unitary evolutions; and (4) T1 and T2 are two tomography sets on the polarization subsystem.}
\end{figure}

In our experiment, we suppose Alice prepares the initial state and sends it to Bob. The state is unknown to Bob, who now can perform local operations on the polarization state to determine the quantum correlations in the initial state. The experimental setup is shown in Fig.~\ref{Fig1}.

On Alice's side, the job is divided into two parts. The first part is the generation of single photons (not shown). The photons are emitted from a single InAs/GaAs QD, which is isolated by a $2$-$\mu$m-diameter pinhole in a gold mask. The exciting power intensity is about $100$ ${\rm{nW}}/\mu{\rm{m}}^{2}$, which makes the QD oversaturated. This power intensity, together with the gold mask on the surface of the QD sample, induces a significant broadening of the QD emission line because of noise effects, such as spectral diffusion \cite{sallen2010,wolters2013}. The frequency state then becomes a mixed state because the inhomogeneous broadening is far larger than the homogeneous broadening. The resulting emission line is then filtered by a Fabry-Perot (FP) cavity with a finesse of 28. The line shape of the single photons is measured by a Michelson interferometer and the result is shown in Fig. 2(a), where $x$ is the delay between the two arms of the interferometer. Assuming a Lorentzian line shape
\begin{equation}
G(\omega)=\frac{\delta\omega}{\pi}\frac{1}{\delta\omega^{2}+(\omega-\omega_{0})^{2}},
\end{equation}
with a center frequency $\omega_0$ corresponding to $914$ nm, we can extract the line width $1/\delta\omega=9.703\pm0.124$ ps, by fitting the measured data to
$\int d\omega G(\omega)e^{-i(\omega-\omega_0) t}=e^{-\delta\omega|t|}$ with $t=2x/c$.

\begin{figure}[tb]
\centering
\includegraphics[width=0.9\linewidth]{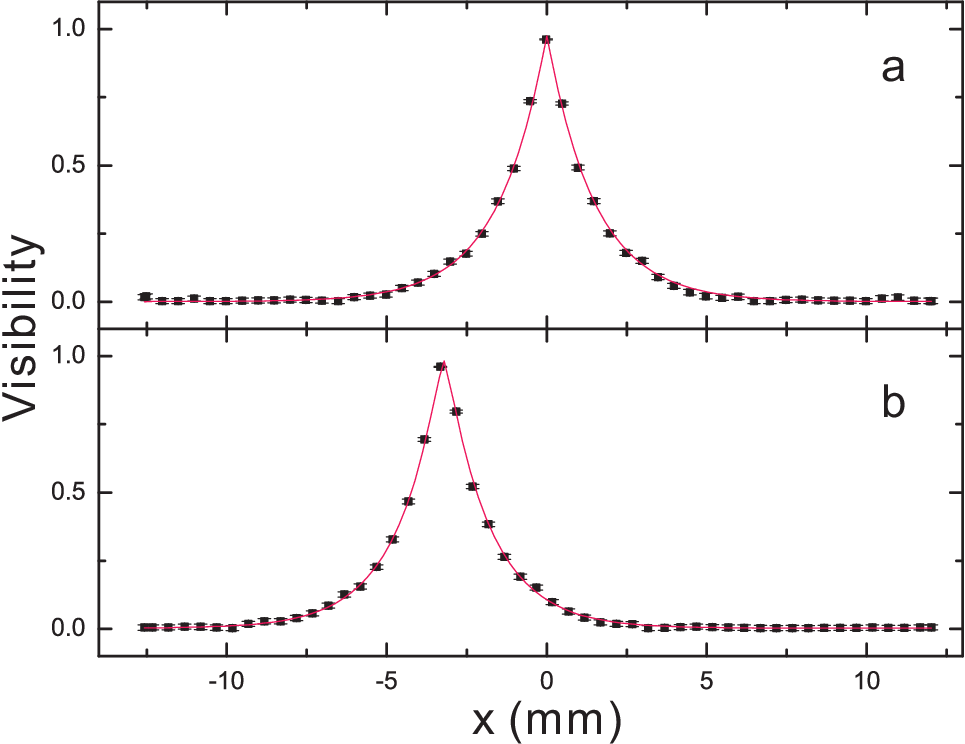}
\caption{\label{Fig2} Visibility of single photons in the Michelson interferometer (a) without and (b) with a calcite, where $x$ is the delay between the two arms. After a Fourier transformation, the data will show the frequency spectrum of the single photons. The red lines show fits based on a Lorentzian spectrum. The shift of the zero point in (b) can be used to calculate the birefringence of the calcite.}
\end{figure}

The second part of Alice's job is the preparation of the initial state $\rho_{SE}$. A $45^{\circ}$ polarizer prepares the photon into the pre-initial state $|H+V\rangle\langle H+V|\otimes\int d\omega G(\omega)|\omega\rangle\langle\omega|$, where $|H\rangle$, $|V\rangle$ are the horizontal and vertical polarization states, respectively. Then, the photon moves through a birefringent quantum channel, generated by a calcite of length $L$ with its optical axis along the vertical direction. Finally a half-wave plate (HWP0) driven by a random number generator (RNG) rotates the basis of the state to a random direction in order to make it impossible
for Bob to know the local eigenbasis on the side of Alice.
At this point, Alice's preparation of the initial state $\rho_{SE}$ is completed and the state is passed on to Bob.

Knowing the parameters used for state preparation, Alice can theoretically determine the quantum correlations in the state $\rho_{SE}$ based on Eq.~(\ref{qctheory}), which in the described situation leads to the following expression (see Supplement 1),
\begin{equation} \label{eq-delta}
 \delta(\rho_{SE}) = \frac{1}{2} \int d\omega \, G(\omega)
 \left| e^{i(\omega - \omega_0)t} - e^{-i(\omega - \omega_0)t} \right|
\end{equation}
with $t=L\Delta n_{\mathrm{cal}}/c$ corresponding to the length $L$ of the calcite.
To measure the birefringence $n_{\mathrm{cal}}$ of the calcite we use again the
Michelson interferometer.
The result for a calcite of length $L=35.92$ mm is shown in Fig.~2(b). From the
shift of the zero delay, we obtain the birefringence
$\Delta n_{\rm{cal}}=0.179\pm0.001$, which coincides with the parameter
provided by the calcite's manufacturer.

The setup on Bob's side consists of four sections. The first section is made up of three removable mirrors (RM1, RM2 and RM3). By placing and removing these mirrors, the photons are led to different functional modules in the experimental setup. One of them is the second section, which implements the controlled dephasing map, Eq.~(\ref{dephasingmap}), using a computer-controlled half-wave plate (HWP1) and a very long polarization-maintaining (PM) optical fiber ($l=120$ m). The third section is a Michelson delay setup, which represents the unitary time evolution. Here, the angle $\eta/2$ of the half-wave plate (HWP2) allows us to generate a set of time-evolution operators $U(\eta,\tau)$. This section further includes a polarizing beam splitter (PBS1), two $45^{\circ}$-placed quarter-wave plates (QWP1 and QWP2), a fixed mirror (M1), and a delayable mirror (M2), whose coordinate $x=c\tau/2$ corresponds to the time variable. The last section is the tomography setup (T1 and T2). T1 includes QWP3, HWP3, PBS2 and two avalanche photodiodes (APD1 and APD2). The setup of T2 is the same as T1.

The measurement of the quantum correlations occurs in two steps. The first step is the preparation of $\rho_{SE}'$. With RM1 in place, the photons are directed to T2, where the tomography of the reduced state $\rho_{S}$ is performed on the polarization subsystem. Then the computer calculates the eigenstates $|\theta\rangle$ and $|\theta_{\perp}\rangle$ of $\rho_{S}$, and controls HWP1 to rotate to the angle $\theta/2$. At this point, we remove RM1 and place RM2, causing the photons to go through the controlled dephasing map. The birefringence of the PM fiber is $\Delta n_{\rm{pmf}}=3\times10^{-4}$, so we have $l\Delta n_{\rm{pmf}}/c\gg1/\delta\omega$. This means that the PM fiber is long enough to completely remove coherences between the orthogonal subspaces spanned by $\{|\theta\rangle\otimes|\omega\rangle\}_{\omega\geq 0}$ and $\{|\theta_{\perp}\rangle\otimes|\omega\rangle\}_{\omega\geq 0}$, which are responsible for the quantum correlations in terms of the quantum discord. This way the state $\rho_{SE}'$ is prepared at the end of the PM fiber.

\begin{figure}[tb]
\centering
\includegraphics[width=0.9\linewidth]{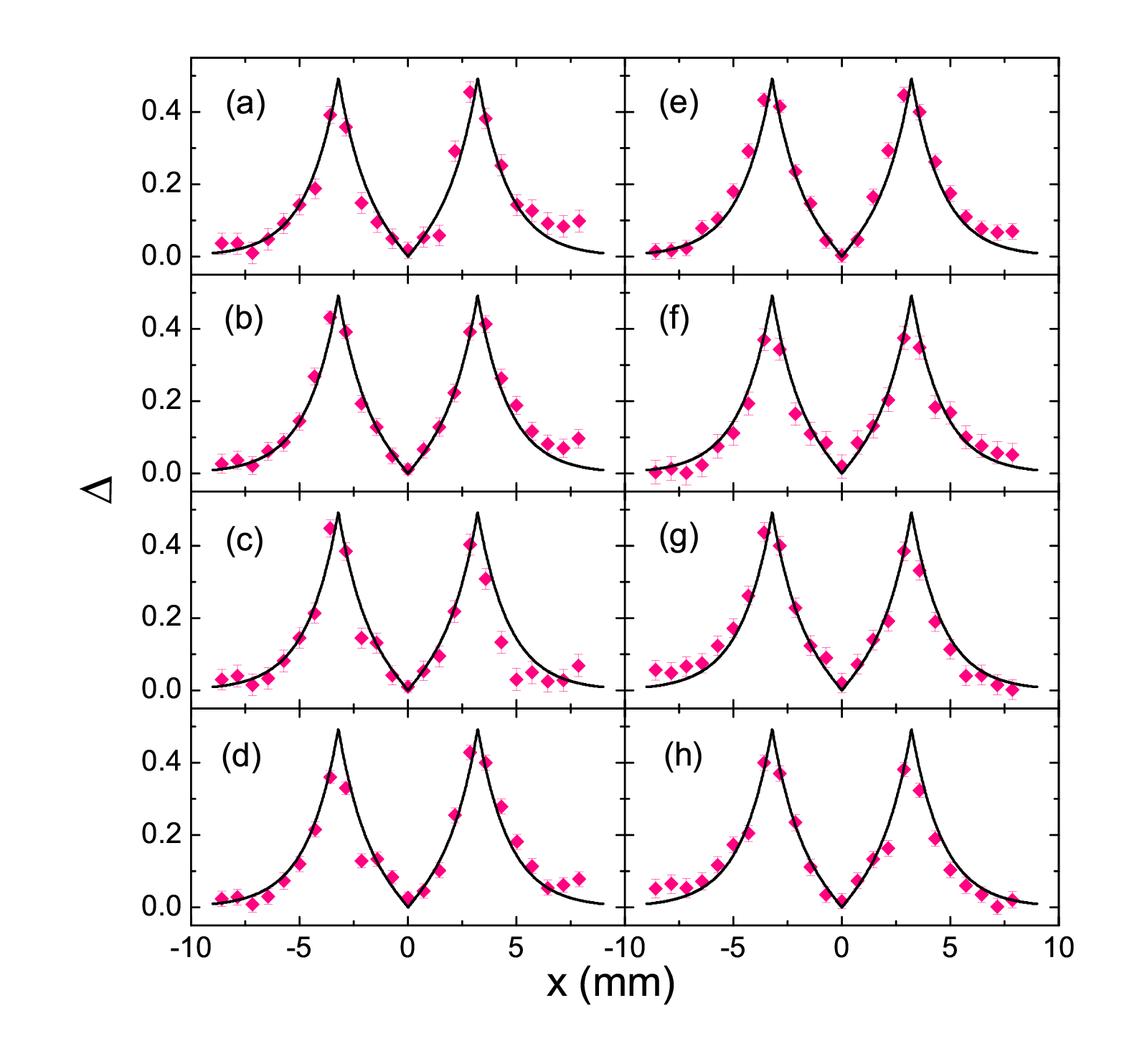}
\caption{\label{Fig3}
Experimental results for the local trace distance $\Delta(\eta,x)$, with $x=c\tau/2$
for different $\eta$-values (from (a) to (h): $\eta=n \pi/16$ with $n=0 \ldots 7$)
and an initial state prepared with $t = L\Delta n_{\mathrm{cal}}/c$ and $L=35.92$
mm. The black lines show the theoretical prediction according to
Eq.~(\ref{eq-Delta-Lorentz}).}
\end{figure}

In the second step, we first remove all of the removable mirrors and send the
photons with state $\rho_{SE}$ to go through the unitary evolution of $U(\eta,\tau)$.
Considering only the polarization state, we obtain the evolved reduced state
$\rho_{S}(\eta,\tau)$. This state can be measured by the tomography setup in T1. We
then place RM2 and RM3. Now, instead of $\rho_{SE}$, we subject $\rho_{SE}'$ to the same evolution, and by local tomography we obtain another reduced state
$\rho_{S}'(\eta,\tau)$. We can then calculate the trace distance
$\Delta(\eta,\tau)=\|\rho_{S}(\eta,\tau)-\rho_{S}'(\eta,\tau)\|$
between the two reduced states.
Here, we choose the parameters $\eta$, $\tau$ as $\eta=m\pi/16$ and
$\tau=(n-12)/2\delta\omega$, with $m=0\dots 7$ and $n=0\dots23$. Taking $L=35.92$ mm as an example, we obtain the results shown in Fig. 3.
The theoretical analysis yields the following expression for the local trace distance (see Supplement 1),
\begin{equation} \label{eq-Delta}
 \Delta(\tau) = \frac{1}{2} \left| \int d\omega G(\omega)
 \left( e^{i(\omega - \omega_0)t} - e^{-i(\omega - \omega_0)t} \right)
 e^{i\omega\tau} \right|.
\end{equation}
Since this expression is independent of $\eta$, as clearly confirmed by the experimental data (see Fig.~\ref{Fig3}), we write $\Delta(\tau)=\Delta(\eta,\tau)$ for simplicity.
Note that according to expressions \eqref{eq-delta} and \eqref{eq-Delta} the general
inequality \eqref{ineq} is, in the present special case, a simple consequence
of the triangular inequality. For the Lorentzian spectrum one finds
\begin{equation} \label{eq-Delta-Lorentz}
 \Delta(\tau) = \frac{1}{2} \left|
 e^{-\delta\omega|t+\tau|} - e^{-\delta\omega|t-\tau|} \right|,
\end{equation}
which is the theoretical prediction shown in Fig.~3.

Now Alice changes the length of the calcite crystal, thus creating a set of different
initial states which correspond to the photon states at different positions in a
quantum channel. The crystal length $L$ is chosen as $L=k\times8.98$ mm
with $k=0\dots6$. For each crystal length, Alice knows the total state and
determines the theoretical quantum correlation $\delta(\rho_{SE})$
employing Eq.~(\ref{eq-delta}). The results are shown in Fig.~4 by the blue line.
Bob receives the initial states, and uses local operations to experimentally detect
the quantum correlations. Using Eq.~(\ref{eq-Delta-Lorentz}) we find that the theoretical
prediction for Bob's measurement results is given by
\begin{equation} \label{eq-max-Delta}
 \max_{\tau} \Delta(\tau) = \frac{1}{2}
 \left( 1 - e^{-2\delta\omega t} \right),
\end{equation}
which is shown by the red line in Fig.~4, and compared to the measurement result shown as black dots. This experimental result is derived by fitting the experimental data of $\Delta(\tau)$ (Fig.~3 shows an example of these data) using Eq.~(\ref{eq-Delta-Lorentz}), and the uncertainty in the result is determined by the fittings. As is evident from the figure, the local
detection scheme of Bob provides a good lower bound for the quantum
correlations in the initial state prepared by Alice: The bound is rather
close to the actual value of the quantum correlations and, moreover, exhibits
the same monotonicity as the initial correlations.

\begin{figure}[tb]
\centering
\includegraphics[width=1\linewidth]{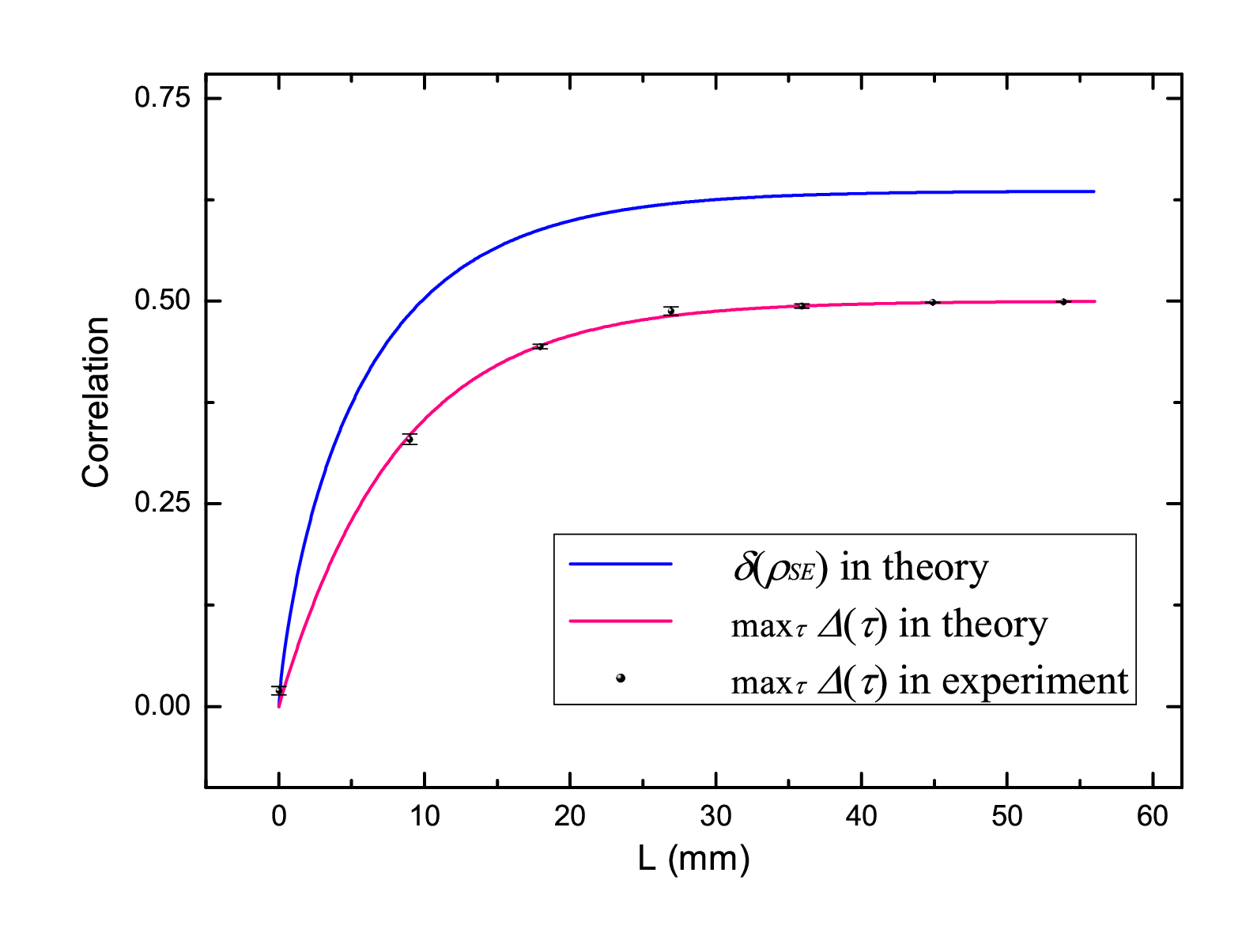}
\caption{\label{Fig4}
Experimental results of the local detection scheme to measure initial quantum
correlations (black dots and error bars). The red line shows the theoretical
prediction given by Eq.~(\ref{eq-max-Delta}), while the blue line represents
the quantum correlations of the initial state according to Eq.~(\ref{eq-delta}).
$L$ is the length of the calcite crystal, which can be considered as the different
positions in a quantum channel.}
\end{figure}

\section{Conclusions}
To conclude, we have prepared quantum correlated initial states in
a photonic system composed of a single photon's polarization state and its
frequency modes using a birefringent quantum channel. Our all-optical experiment
demonstrates that these quantum correlations can be detected and quantified by
performing only local operations on the polarization subsystem. The theoretical
analysis shows that the local detection scheme leads to a close lower bound for
the quantum correlations of the total initial state. The ability to detect and estimate these bipartite correlations with access to only one of the subsystems distinguishes our work from previously developed methods \cite{Girolami2012}, which had been realized experimentally with two-qubit systems based on nuclear magnetic resonance \cite{Silva2013,Paula2013}. The methods presented in Refs. \cite{Girolami2012,Silva2013} allow for estimation of the correlations of unknown states by local operations on \textit{both} subsystems if one of the subsystems is a qubit, and both subsystems are finite-dimensional and experimentally under control. In contrast, the local detection method employed here only requires control over one of the two systems (of arbitrary dimensions) and the presence of a suitable interaction between them.

Finally, we emphasize that our work represents the first experimental realization of this scheme for mixed
correlated initial states in an infinite-dimensional tensor product Hilbert space,
where a finite-dimensional truncation is prevented by the contributions of a truly
infinite number of orthogonal continuous-variable states. The developed
experimental techniques can be naturally extended to the study of correlations in
other photonic open systems and quantum channels for which only a local part
can be controlled and measured.

\section*{Funding Information}
This work is supported by the National Natural Science Foundation of China (Grants Nos. 61490711, 11274289, 11304305, 11325419, 61327901 and 91321313), the National Basic Research Program of China (Grants Nos. 2011CB921200 and 2013CB933304), the Strategic Priority Research Program(B) of the Chinese Academy of Sciences (Grant No. XDB01030300), the Fundamental Research Funds for the Central Universities (No. WK2470000011), the China Postdoctoral Science Foundation funded project (Grant No. 2012M521229), and the German Academic Exchange
Service (DAAD). M.G. thanks the German National Academic Foundation for support. C.-F.L. and H.-P.B. acknowledge support from the European Union (EU) through the Collaborative Project QuProCS (Grant
Agreement No. 641277).

\section*{}
See Supplement 1 for supporting content.

\newpage
\begin{center}
\textbf{SUPPLEMENTARY MATERIAL}
\end{center}

\section*{}
\subsection{Preparation of the initial state}
\subsubsection{The pre-initial state}
The pre-initial state is a product state, $\rho_{\rm{pi}}=\rho_0\otimes\rho_E$, where the density matrix of the two-level system represented in the $\{|H\rangle,|V\rangle\}$ basis reads
\begin{align}
\rho_0=\begin{pmatrix}1/2 & de^{i\varphi}\\de^{-i\varphi} & 1/2\end{pmatrix},\tag{S1}
\end{align}
and the environmental state is a mixed state given by a sum over closely spaced frequencies,
\begin{align}
\rho_E=\sum_{\omega}\Delta\omega G(\omega)|\omega\rangle\langle\omega|.\tag{S2}
\end{align}
In order to avoid problems with non-normalizable continuum states, we formally
represent the state $\rho_E$ by a discrete sum of frequencies, and perform
the continuum limit
($\Delta\omega\rightarrow 0$) after obtaining the relevant quantities, such as
the trace distance. The frequency states $|\omega\rangle$ form a complete
orthonormal set, i.e., $\langle\omega|\omega'\rangle=\delta_{\omega,\omega'}$
and $\sum_{\omega}|\omega\rangle\langle\omega|=\mathbb{I}$. The conditions
$\Delta\omega G(\omega)\geq0$ and
$\sum_{\omega}\Delta\omega G(\omega)=1$ then ensure that $\rho_E$ indeed
describes a positive and normalized quantum state. In the following we assume
for simplicity that $G(\omega)$ is a unimodal function with a width of order
$\delta\omega$ which is symmetric about the center frequency $\omega_0$.

\subsubsection{The correlated initial state}
In order to prepare a quantum state with polarization-frequency correlations, we send the pre-initial state
\begin{align}
\rho_{\rm{pi}}&=\sum_{\omega}\Delta\omega G(\omega)\left(\frac{1}{2}|H,\omega\rangle\langle H,\omega| +de^{i\varphi}|H,\omega\rangle\langle V,\omega| \right.\notag\\
&\left.\hspace{1.2cm}+\:de^{-i\varphi}|V,\omega\rangle\langle H,\omega|+\frac{1}{2}|V,\omega\rangle\langle V,\omega|\right)\tag{S3}
\end{align}
through a birefringent crystal, which generates an interaction of the form
\begin{align}
U_{\rm{cal}}(t):\begin{cases}|H,\omega\rangle\rightarrow|H,\omega\rangle\\
|V,\omega\rangle\rightarrow e^{-i\omega t}|V,\omega\rangle\end{cases},\tag{S4}
\end{align}
where $t=\Delta_{\rm{cal}} l/c$ corresponds to the time spent inside the crystal. The state $\rho_{\rm{pi}}$ evolves into the initial state
\begin{align}\label{eq.initstate}
&\rho_{SE}\notag\\
&=U_{\rm{cal}}(t)\rho_{\rm{pi}}U^{\dagger}_{\rm{cal}}(t)\notag\\
&=\sum_{\omega}\Delta\omega G(\omega)\left(\frac{1}{2}|H,\omega\rangle\langle H,\omega|+de^{i(\omega t+\varphi)}|H,\omega\rangle\langle V,\omega|\right.\notag\\
&\hspace{1.3cm}\left.+\:de^{-i(\omega t+\varphi)}|V,\omega\rangle\langle H,\omega|+\frac{1}{2}|V,\omega\rangle\langle V,\omega|\right).\tag{S5}
\end{align}
Its reduced state for the two-level system is given by
\begin{align}\label{eq.reducedstate}
\rho_S=\begin{pmatrix}1/2 & dC(t)e^{i\Psi(t)}\\
dC(t)e^{-i\Psi(t)}&1/2\end{pmatrix},\tag{S6}
\end{align}
with $\Psi(t)=\omega_0t+\varphi$ and $\omega_0$ denotes the center frequency of the distribution $G(\omega)$. For fixed $t$, the initial phase $\varphi$ can be chosen such that $\Psi(t)\equiv 0$, i.e., $\varphi=-\omega_0t$. Since $G(\omega)$ is symmetric about $\omega=\omega_0$, we obtain a real-valued function
\begin{align}
C(t)=\sum_{\omega}\Delta\omega G(\omega)e^{i(\omega-\omega_0)t}\tag{S7}
\end{align}

For the implementation of the local detection protocol, the first step consists in obtaining the eigenbasis of $\rho_S$. Diagonalization of this $2\times2$ matrix yields the eigenvalues,
\begin{align}
\lambda_0&=1/2+dC(t),\notag\\
\lambda_1&=1/2-dC(t),\tag{S8}
\end{align}
and corresponding eigenvectors,
\begin{align}\label{eq.eigenbasis}
|0\rangle&=\frac{1}{\sqrt{2}}\begin{pmatrix}1 \\ 1\end{pmatrix}=\frac{1}{\sqrt{2}}(|H\rangle+|V\rangle),\notag\\
|1\rangle&=\frac{1}{\sqrt{2}}\begin{pmatrix}1 \\ -1\end{pmatrix}=\frac{1}{\sqrt{2}}(|H\rangle-|V\rangle).\tag{S9}
\end{align}
Before passing the state on to Bob, Alice implements a random local unitary rotation $U_r$ to hide her local eigenbasis. This is implemented by a half-wave plate which is controlled by a random number generator. As a consequence, the eigenstates are rotated to $|\theta\rangle=U_r|0\rangle$ and $|\theta_{\perp}\rangle=U_r|1\rangle$. This local unitary operation has no effect on the correlation properties of the state. Redefining the basis vectors $|H\rangle$ and $|V\rangle$ accordingly always allows us to map the state back onto the original non-rotated state. Thus, for the following theoretical analysis, we can restrict to the case $U_r=\mathbb{I}$, i.e., effectively disregard this additional local rotation.

\subsection{The local dephasing operation}
The local dephasing must be carried out in the eigenbasis of the reduced state (\ref{eq.reducedstate}). Hence, the desired operation is
\begin{align}
\Phi(X)=|0\rangle\langle0|X|0\rangle\langle0|+|1\rangle\langle1|X|1\rangle\langle1|.\tag{S10}
\end{align}
Local application to the state $\rho_{SE}$ yields
\begin{align}
\rho'_{SE}&=(\Phi\otimes\mathbb{I})\rho_{SE}\tag{S11}\\
&=\frac{1}{2}\sum_{\omega}\Delta\omega G(\omega)\Big[|H,\omega\rangle\langle H,\omega|+|V,\omega\rangle\langle V,\omega|\notag\\
&\qquad+\:d(e^{i(\omega-\omega_0) t}+e^{-i(\omega-\omega_0) t})|H,\omega\rangle\langle V,\omega|\notag\\
&\qquad+\:d(e^{i(\omega-\omega_0) t}+e^{-i(\omega-\omega_0) t})|V,\omega\rangle\langle H,\omega|\Big],\notag
\end{align}
where we have used Eq.~(\ref{eq.eigenbasis}).
\subsubsection{Experimental implementation}
We implement the controlled dephasing operation by sending the quantum state through a very long polarization-maintaining fiber after rotating the eigenstates $|0\rangle$ and $|1\rangle$ onto the principal axes of the fiber with a computer-controlled half-wave plate. Formally, the generated evolution is then described by
\begin{equation}
 U_{\rm deph}: \left\{
 \begin{array}{l}
 |0,\omega\rangle \rightarrow |0,\omega\rangle \\
 |1,\omega\rangle \rightarrow e^{-i\omega s} |1,\omega\rangle
 \end{array} \right. ,\tag{S12}
\end{equation}
where $s=\Delta_{\rm{pmf}} l/c \gg 1/\delta\omega$ corresponds to the time the photon spends inside the fiber. This can be reformulated as
\begin{equation}
 U_{\rm deph}: \left\{
 \begin{array}{l}
 |H,\omega\rangle \rightarrow a|H,\omega\rangle + b|V,\omega\rangle\\
 |V,\omega\rangle \rightarrow b|H,\omega\rangle + a|V,\omega\rangle
 \end{array} \right.\tag{S13}
\end{equation}
with
\begin{equation}
 a = \frac{1}{2}\left( 1 + e^{-i\omega s} \right), \;\;\;
 b = \frac{1}{2}\left( 1 - e^{-i\omega s} \right).\tag{S14}
\end{equation}
We thus obtain
\begin{align}
 &\rho''_{SE} \tag{S15}\\
 =\:& U_{\rm deph} \rho_{SE} U^{\dagger}_{\rm deph}\notag \\
 =\:&
  \sum_{\omega} \Delta\omega G(\omega) \left[
 \frac{1}{2}|H,\omega\rangle\langle H,\omega| + \frac{1}{2}|V,\omega\rangle\langle V,\omega| \right. \notag\\
 &
 + d \left( e^{i(\omega-\omega_0)t}ab^* + e^{-i(\omega-\omega_0)t}a^*b \right)
 |H,\omega\rangle\langle H,\omega| \notag \\
 &
 + d \left( e^{i(\omega-\omega_0)t}a^*b + e^{-i(\omega-\omega_0)t}ab^* \right)
 |V,\omega\rangle\langle V,\omega|\notag  \\
 & + d \left( e^{i(\omega-\omega_0)t}|a|^2 + e^{-i(\omega-\omega_0)t}|b|^2 \right)
 |H,\omega\rangle\langle V,\omega| \notag \\
 & \left.
 + d \left( e^{i(\omega-\omega_0)t}|b|^2 + e^{-i(\omega-\omega_0)t}|a|^2 \right)
 |V,\omega\rangle\langle H,\omega|\notag
 \right].
\end{align}
This state realizes the desired dephased state $\rho'_{SE}$ to a good approximation. To see this, consider for instance,
\begin{equation}
 |a|^2 = \frac{1}{2} ( 1 + \cos\omega s ).\tag{S16}
\end{equation}
Due to the condition $\delta\omega \cdot s \gg 1$, the function
$f(\omega)=\cos\omega s $ oscillates very rapidly compared to variations of
$G(\omega)$ and the remaining exponential functions. Such rapid oscillations
cannot be resolved and, hence, $|a|^2$ must be replaced by its mean value over
one period (in frequency space):
\begin{equation}
 |a|^2 \rightarrow \langle |a|^2 \rangle = \frac{1}{2}.\tag{S17}
\end{equation}
We obtain correspondingly
\begin{equation}
 |b|^2 \rightarrow \langle |b|^2 \rangle = \frac{1}{2}, \;\;\;
 ab^* \rightarrow \langle ab^* \rangle = 0.\tag{S18}
\end{equation}
With these replacements $\rho''_{SE}$ reduces to $\rho'_{SE}$.
\subsubsection{Reduced states}
The reduced states of $\rho_{SE}$ and $\rho'_{SE}$ coincide by construction, i.e., $\mathrm{Tr}_E\rho_{SE}=\mathrm{Tr}_E\rho'_{SE}$ and $\mathrm{Tr}_S\rho_{SE}=\mathrm{Tr}_S\rho'_{SE}$. This can be confirmed explicitly as follows. First note that
\begin{align}
&\quad\rho_{SE}-\rho'_{SE}\notag\\&=\sum_{\omega}\Delta\omega G(\omega) f(\omega)\left[|H,\omega\rangle\langle V,\omega|-|V,\omega\rangle\langle H,\omega|\right],\tag{S19}
\end{align}
where
\begin{align}
f(\omega)=\frac{d}{2}\left(e^{i(\omega-\omega_0) t}-e^{-i(\omega-\omega_0) t}\right)=id\sin\left[(\omega-\omega_0) t\right]\tag{S20}
\end{align}
is purely imaginary and anti-symmetric about $\omega=\omega_0$. One immediately confirms that
\begin{align}
\rho_E-\rho'_E=\mathrm{Tr}_S\{\rho_{SE}-\rho'_{SE}\}=0.\tag{S21}
\end{align}
Tracing over the environmental degrees of freedom thus yields
\begin{align}
&\qquad\rho_S-\rho'_S\notag\\&=\mathrm{Tr}_E\{\rho_{SE}-\rho'_{SE}\}\notag\\&=\sum_{\omega}\Delta\omega G(\omega) f(\omega)\left[|H\rangle\langle V|-|V\rangle\langle H|\right].\tag{S22}
\end{align}
We now perform the continuum limit $\Delta\omega\rightarrow0$, leading to
\begin{align}
\sum_{\omega}\Delta\omega G(\omega) f(\omega)\longrightarrow\int d\omega G(\omega)f(\omega)=0.\tag{S23}
\end{align}
In the final step, we have used that $G(\omega)$ is a symmetric function about $\omega=\omega_0$, while $f(\omega)$ is anti-symmetric. Hence, we confirm
\begin{align}
\rho_S-\rho'_S=0.\tag{S24}
\end{align}
\subsubsection{Total trace distance}
The difference between the two states $\rho_{SE}$ and $\rho'_{SE}$ is given by
\begin{align}
X&=\rho_{SE}-\rho'_{SE}\notag\\
&=\frac{1}{2}\sum_{\omega} \Delta\omega G(\omega)d(e^{i(\omega-\omega_0) t}-e^{-i(\omega-\omega_0) t})\notag\\
&\hspace{1.5cm}\times\left(|H,\omega\rangle\langle V,\omega|-|V,\omega\rangle\langle H,\omega|\right).\tag{S25}
\end{align}
Diagonalization yields eigenvalues
\begin{align}
\mu_{\pm}(\omega)=\pm i\Delta\omega G(\omega)f(\omega)\tag{S26}
\end{align}
and eigenstates
\begin{align}
|\mu_{\pm}(\omega)\rangle=\frac{1}{\sqrt{2}}(|H\rangle\pm i|V\rangle)\otimes|\omega\rangle.\tag{S27}
\end{align}
The trace norm of $X$, which corresponds to the trace distance of $\rho_{SE}$ and $\rho'_{SE}$ can thus be written as
\begin{align}
\delta(\rho_{SE})&=\|\rho_{SE}-\rho'_{SE}\|\notag\\
&=\mathrm{Tr}\sqrt{X^{\dagger}X}\notag\\
&=\sum_{\omega}\left(|\mu_-(\omega)|+|\mu_+(\omega)|\right)\notag\\
&=2\sum_{\omega}|\Delta\omega G(\omega)f(\omega)|\notag\\
&=2\sum_{\omega}\Delta\omega G(\omega)|f(\omega)|\notag\\
&=d\sum_{\omega}\Delta\omega G(\omega)\left|e^{i(\omega-\omega_0) t}-e^{-i(\omega-\omega_0) t}\right|.\tag{S28}
\end{align}
Performing the continuum limit finally leads to
\begin{align}\label{eq.totaldistance}
\delta(\rho_{SE})=d\int d\omega G(\omega)\left|e^{i(\omega-\omega_0) t}-e^{-i(\omega-\omega_0) t}\right|,\tag{S29}
\end{align}
which for the parameter used in the experiment ($d=1/2$) yields the result given in Eq.~(7) of the main manuscript.

\subsection{Open-system evolution depending on initial correlations}
For the dynamical local detection of the correlations in $\rho_{SE}$, we subject both $\rho_{SE}$ and $\rho'_{SE}$ to the Michelson delay setup (see main manuscript), generating the dynamics
\begin{align}
U(\eta,\tau):\begin{cases}|\eta,\omega\rangle&\longrightarrow|\eta,\omega\rangle\\
|\eta_{\perp},\omega\rangle&\longrightarrow e^{-i\omega \tau}|\eta_{\perp},\omega\rangle\end{cases},\tag{S30}
\end{align}
where $\tau=2x/c$ correponds to the coordinate of the delayable mirror (M2) and the half-wave plate (HWP2) is rotated to the angle $\eta/2$, generating the basis vectors
\begin{align}
|\eta\rangle&=\begin{pmatrix}\cos\eta \\\sin\eta\end{pmatrix},\notag\\
|\eta_{\perp}\rangle&=\begin{pmatrix}-\sin\eta \\\cos\eta\end{pmatrix}.\tag{S31}
\end{align}
The original polarization vectors $|H\rangle$ and $|V\rangle$ can be expressed in terms of the new basis vectors $|\eta\rangle$ and $|\eta_{\perp}\rangle$ as
\begin{align}
|H\rangle&=\cos\eta|\eta\rangle-\sin\eta|\eta_{\perp}\rangle,\notag\\
|V\rangle&=\sin\eta|\eta\rangle+\cos\eta|\eta_{\perp}\rangle.\tag{S32}
\end{align}
Hence, the evolution of the original states is given by
\begin{align}
U(\eta,\tau):\begin{cases}|H,\omega\rangle&\longrightarrow\cos\eta|\eta,\omega\rangle-\sin\eta e^{-i\omega\tau}|\eta_{\perp},\omega\rangle\\
|V,\omega\rangle&\longrightarrow\sin\eta|\eta,\omega\rangle+\cos\eta e^{-i\omega \tau}|\eta_{\perp},\omega\rangle\end{cases}.\tag{S33}
\end{align}

\subsubsection{Local trace distance}
The operator $X=\rho_{SE}-\rho'_{SE}$ evolves under this dynamics as
\begin{align}
&\quad U(\eta,\tau)(\rho_{SE}-\rho'_{SE})U^{\dagger}(\eta,\tau)\notag\\&=\sum_{\omega}\Delta\omega G(\omega)f(\omega)\notag\\&\quad\times\left[\sin\eta\cos\eta|\eta,\omega\rangle\langle\eta,\omega|+\cos^2\eta e^{i\omega\tau}|\eta,\omega\rangle\langle\eta_{\perp},\omega|\notag\right.\\&\qquad-\sin^2\eta e^{-i\omega\tau}|\eta_{\perp},\omega\rangle\langle\eta,\omega|-\sin\eta\cos\eta|\eta_{\perp},\omega\rangle\langle\eta_{\perp},\omega|\notag\\&\qquad\left.-\mathrm{H.c.}\right]\notag\\
&=\sum_{\omega}\Delta\omega G(\omega)f(\omega)\notag\\&\quad\times\left[e^{i\omega\tau}|\eta,\omega\rangle\langle\eta_{\perp},\omega|-e^{-i\omega\tau}|\eta_{\perp},\omega\rangle\langle\eta,\omega|\right].\tag{S34}
\end{align}
Since we only observe the dynamics of the polarization states, we take the partial trace, which yields
\begin{align}
&\quad\mathrm{Tr}_E\{U(\eta,\tau)(\rho_{SE}-\rho'_{SE})U^{\dagger}(\eta,\tau)\}\notag\\&=\sum_{\omega}\Delta\omega G(\omega)f(\omega)\left[e^{i\omega\tau}|\eta\rangle\langle\eta_{\perp}|-e^{-i\omega\tau}|\eta_{\perp}\rangle\langle\eta|\right].\tag{S35}
\end{align}
Represented as a matrix in the basis $\{|\eta\rangle,|\eta_{\perp}\rangle\}$, this reads
\begin{align}\label{eq.localtracedistoperator}
\mathrm{Tr}_E\{U(\eta,\tau)(\rho_{SE}-\rho'_{SE})U^{\dagger}(\eta,\tau)\}=\begin{pmatrix} 0 & z \\ z^* & 0\end{pmatrix},\tag{S36}
\end{align}
where
\begin{align}
z=\sum_{\omega}\Delta\omega G(\omega)f(\omega)e^{i\omega\tau}\tag{S37}
\end{align}
Note that this result is independent of $\eta$. The trace norm of the operator~(\ref{eq.localtracedistoperator}) corresponds to the trace distance of the locally accessible polarization states $\rho_S(\tau)$ and $\rho'_S(\tau)$ and can be determined as
\begin{align}
\Delta(\tau)&=\|\rho_S(\tau)-\rho'_S(\tau)\|\tag{S38}\\
&=2|z|\notag\\
&=2\left|\sum_{\omega}\Delta\omega G(\omega)f(\omega)e^{i\omega\tau}\right|\notag\\
&=d \left|\sum_{\omega}\Delta\omega G(\omega)
 \left( e^{i(\omega-\omega_0) t} - e^{-i(\omega-\omega_0) t} \right) e^{i\omega\tau}
 \right|.\notag
\end{align}
The continuum limit then yields
\begin{align}\label{eq.localdistance}
\Delta(\tau)&=d \left|\int d\omega G(\omega)
 \left( e^{i(\omega-\omega_0) t} - e^{-i(\omega-\omega_0) t} \right) e^{i\omega\tau}
 \right|.\tag{S39}
\end{align}
By comparing this result to Eq.~(\ref{eq.totaldistance}), one readily confirms that the general inequality
\begin{align}
\|\mathrm{Tr}_E\{U(\eta,\tau)(\rho_{SE}-\rho'_{SE})U^{\dagger}(\eta,\tau)\}\|\leq\|\rho_{SE}-\rho'_{SE}\|\tag{S40}
\end{align}
here is satisfied as a consequence of the triangular inequality.

In the present experiment, the frequency distribution of $\rho_E$ is characterized by a Lorentzian spectrum
\begin{align}
G(\omega)=\frac{1}{\pi}\frac{\delta\omega}{\delta\omega^2+(\omega-\omega_0)^2}.\tag{S41}
\end{align}
We have $\omega_0=2\pi c/(914\,\mathrm{nm})$, and
$1/\delta\omega = 9.703 \pm 0.124$ ps. The function $G(\omega)$ is normalized and has its support to a good approximation in the positive domain since $\omega_0\gg\delta\omega>0$,
\begin{align}
\int_{-\infty}^{\infty}G(\omega)\simeq\int_{0}^{\infty}G(\omega)=1.\tag{S42}
\end{align}
Its Fourier transform is given by
\begin{align}\label{eq.ft}
\int_{0}^{\infty}G(\omega)e^{\pm i\omega t}d\omega=e^{\pm i\omega_0t}C(t),\tag{S43}
\end{align}
with
\begin{align}
C(t)=e^{-\delta\omega |t|}.\tag{S44}
\end{align}
With this, the local trace distance, Eq.~(\ref{eq.localdistance}), reduces to
\begin{align}
\Delta(\tau)=d\left|e^{-\delta\omega|t+\tau|}-e^{-\delta\omega|t-\tau|}\right|,\tag{S45}
\end{align}
which for $d=1/2$ yields Eq.~(9) of the main manuscript.

\end{document}